\documentclass[%
 reprint,
superscriptaddress,
 amsmath,amssymb,
 aps,
]{revtex4-1}

\usepackage{graphicx}
\usepackage{dcolumn}
\usepackage{bm}

\newcommand{\D}{\ensuremath{\mathrm{d}}}
\newcommand{\diag}{\ensuremath{\operatorname{diag}}}
\newcommand{\Tr}{\ensuremath{\operatorname{Tr}}}

\begin{document}

\title{Comparing molecules and solids across structural and alchemical space}
\author{Sandip De}
\affiliation{National Center for Computational Design and Discovery of Novel Materials (MARVEL)}
\affiliation{Laboratory of Computational Science and Modelling, Institute of Materials, Ecole Polytechnique F\'ed\'erale de Lausanne, Lausanne, Switzerland}
\author{Albert P. Bart{\'o}k}
\affiliation{Engineering Laboratory, University of Cambridge, Trumpington Street, Cambridge CB2 1PZ, United Kingdom}
\author{G\'abor Cs\'anyi}
\affiliation{Engineering Laboratory, University of Cambridge, Trumpington Street, Cambridge CB2 1PZ, United Kingdom}
\author{Michele Ceriotti}
\affiliation{National Center for Computational Design and Discovery of Novel Materials (MARVEL)}
\affiliation{Laboratory of Computational Science and Modelling, Institute of Materials, Ecole Polytechnique F\'ed\'erale de Lausanne, Lausanne, Switzerland}
\begin{abstract}
Evaluating the (dis)similarity of crystalline, disordered and
molecular compounds is a critical step in the development 
of algorithms to navigate automatically the configuration
space of complex materials. For instance, a structural 
similarity metric is crucial for classifying structures, searching 
chemical space for better compounds and materials, and driving the next
generation of machine-learning techniques for predicting
the stability and properties of molecules and materials.  In the last
few years several strategies have been designed to compare
atomic coordination environments. In particular, the Smooth
Overlap of Atomic Positions (SOAP) has emerged as an elegant 
framework to obtain translation, rotation and permutation-invariant
descriptors of groups of atoms, driven by
the design of various classes of machine-learned
inter-atomic potentials. Here we discuss how one can combine
such local descriptors using a Regularized Entropy Match (REMatch)
approach to describe the similarity of  
both whole molecular and bulk periodic structures,
introducing powerful metrics that enable the navigation of  
alchemical and structural complexity within a unified framework.
Furthermore, using this kernel and a ridge regression method we can 
predict atomization energies for a database of small organic molecules
with a mean absolute error below 1kcal/mol, reaching an important
milestone in the application of machine-learning techniques
to the evaluation of molecular properties.
\end{abstract}

\maketitle

\section{\label{sec:intro}Introduction}
 
The increase of available computational power, together with 
the development of more accurate and efficient simulation algorithms,
have made it possible to reliably predict  the properties of materials
and molecules of increasing levels of complexity. 
Furthermore, high-throughput computational screening of existing
and hypothetical compounds promises to dramatically accelerate 
the development of materials with the better performances or
custom-tailored properties \cite{PhysRevLett.114.105503,
PhysRevB.92.014106,PhysRevB.92.094306,kusne15screport,
ramkrsinan_2014sd,PhysRevB.90.155136}. 

These developments have made even more urgent the need
for automated tools to analyze, classify~\cite{rodr-laio14science,
clustering-rui,gang_clustering,cartography,prasanna-sd}
and represent~\cite{ferg+10pnas,ceri+11pnas,trib+12pnas,ceri+13jctc,rohr+arpc13}
large amounts of structural data, as well as techniques to
leverage this wealth of information to estimate inexpensively
the properties of materials using machine-learning techniques,
circumventing the need for computationally demanding
quantum mechanical calculations \cite{PhysRevLett.108.058301,
PhysRevB.90.104108,PhysRevB.89.235411,
pilania13screport,PhysRevB.88.054104,rupp+07jcim,
hirn+15arxiv,qm7b,PhysRevLett.108.253002,PhysRevB.92.045131,anatoleIJQC,bag.of.bonds}.

At the most fundamental level, the crucial ingredient for
all these techniques is a mathematical formulation of the 
concept of (dis)similarity between atomic configurations, 
that can take the form of a distance - that can be used 
for dimensionality reduction or clustering - 
or of a kernel function, that could be used for 
ridge regression or automated classification.\cite{krr-plant,krr-face,rasm05book,hast09book}
The most obvious choice for a metric to compare atomic structures
would involve the Euclidean distance between the Cartesian
coordinates of the atoms, commonly known as root mean square
displacement (RMSD) distance, that can be easily made
invariant to relative translations and rotations. It 
is however highly non-trivial to extend the RMSD to 
deal with situations in which atoms in the two structures
cannot be mapped unequivocally onto each other. The 
deterministic evaluation of a ``permutationally invariant''
RMSD scales combinatorially with the size of the molecules
to be compared~\cite{sade+13jcp}, and introduces cusps at locations where the mapping of atom identities changes. Furthermore, as we will 
discuss later on, the RMSD is perhaps the most straightforward,
but not necessarily the most flexible or effective strategy
to compare molecular and condensed-phase configurations.

In the last few years, a large number of ``fingerprint''
functions have been developed to represent the state of 
structures, or of groups of atoms within a structure.
Structural descriptors have been developed based on graph-theoretic
procedures (e.g. SPRINTs~\cite{sprint}), as well as on 
analogies with electronic structrure methods (e.g. 
Hamiltonian matrix, Hessian matrix, Overlap matrix of Gaussian 
type Orbitals (GTO) or even Kohn-Sham eigenvalues fingerprints~\cite{sade+13jcp}). 
Most of these approaches have been introduced to provide a fast and 
reliable estimate of the dissimilarity between structures. 
Several other descriptors have been also used in machine learning,   
to predict properties of materials and molecules circumventing 
the need for an expensive electronic structure calculation. 
A non-comprehensive list of such methods include Coulomb 
matrices~\cite{PhysRevLett.108.058301}, bags of bonds~\cite{bag.of.bonds}, 
``symmetry functions''~\cite{behlerNNP}, scattering transformation 
applied on a linear superposition of atomic densities~\cite{hirn+15arxiv}.

A particularly promising approach to compare structures
in a way that is invariant to rotations, translations,
and permutations of equivalent atoms, is to start from 
descriptors designed to represent \emph{local} atomic
environments and that fulfill these requirements, and 
combine them to yield a \emph{global} measure of 
similarity between structures. This idea typically relies
on finding the best match between pairs of environments
in the two configurations~\cite{rupp+07jcim,sade+13jcp,newstefan}, and 
can also be traced back to methods developed to compare
images based on the matching of local features~\cite{grau05ieee}.

In the present work we start from a recently-developed
strategy to define a similarity kernel between local
environments -- the smooth overlap of atomic positions 
(SOAP)\cite{bart+13prb} -- and discuss the different ways one 
can process the set of all possible matchings between atomic environments
to generate a global kernel to compare two structures.
In particular, we introduce a regularized entropy match
(REMatch) strategy that is based on
techniques in optimal-transport 
theory~\cite{cutu13nips}, and that is both more efficient
and tunable than previously-applied methods.
We discuss the relative merits of different approaches, 
and generalize this strategy to the comparison between 
structures with different numbers and kinds of atoms.
We demonstrate the behavior of the different global
kernels when applied to completely different classes of problems,
ranging from elemental clusters, to bulk structures, to 
the conformers of oligopeptides and to a heterogeneous
database of small organic molecules. We visualize 
the behavior of the distance associated with these kernels
using sketch-map~\cite{ceri+11pnas}, a non-linear dimensionality reduction
technique, and demonstrate the great promise shown
by the straightforward application of the REMatch-SOAP
kernel to the machine-learning of molecular properties.
Finally, we present our conclusions.

\section{\label{sec:theory}Theory}

Let us start by introducing the notation we will employ in the
rest of the paper. We will label structures to be compared by 
capital letters, use a lowercase Latin letter to indicate 
the index of an atom, and when necessary use a 
Greek lowercase letter to mark its chemical identity.
For instance, the position of the $i$-th atom 
within the structure $A$ will be labeled as $\mathbf{x}^A_{i}$. 
The \emph{environment} of that atom, i.e. the abstract descriptor of the
arrangement of atoms in its vicinity will be labelled with 
a calligraphic upper case letter, e.g. $\mathcal{X}^A_{i}$, 
and the sub-set of such environment that singles out atoms of
species $\alpha$ will be indicated as $\mathcal{X}^{A,\alpha}_{i}$.

Among the many descriptors of local environments that have been
developed in the recent years\cite{PhysRevLett.114.105503,
PhysRevB.92.014106,PhysRevB.92.094306,ramkrsinan_2014sd,
PhysRevB.90.155136,rupp+07jcim,sade+13jcp,
newstefan,PhysRevLett.108.058301,PhysRevB.90.104108,
PhysRevB.89.235411,pilania13screport,PhysRevB.88.054104,qm7b,
PhysRevLett.108.253002,PhysRevB.92.045131,anatoleIJQC,bag.of.bonds}, 
we will refer in particular to the SOAP fingerprints \cite{bart+13prb},  that have
been proven to be a very elegant and robust strategy to describe
coordination environments in a way that is naturally invariant 
with respect to translations, rotations and permutations of atoms. 
We will use the notation $k(\mathcal{X},\mathcal{X}')$ to
indicate the similarity kernel (normalized to one) 
between two environments 
-- which one would use in a kernel ridge regression method~\cite{hast09book,scho+98nc,rasm05book} --
and $d(\mathcal{X},\mathcal{X}')^2=2-2k(\mathcal{X},\mathcal{X}')$
to indicate the (squared) kernel distance between the environments 
-- which one would use in a dimensionality reduction 
method~\cite{ceri+11pnas,rohr+arpc13}.
In what follows we will discuss different ways by which 
environment kernels can be combined to yield a 
a \emph{global} similarity kernel between two structures $K(A,B)$,
and the associated squared distance $D(A,B)^2=2-2K(A,B)$.

\subsection{SOAP similarity kernels and local environment distance}

We will first focus on the comparison between the environment 
of two atoms in a pure compound made up of a single atomic species $\alpha$. 
The crucial ingredient in making the comparison is a kernel
function based on the distribution of atoms in the two 
environments. In the context of SOAP kernels
one represents the local density of atoms within the 
environment $\mathcal{X}$ as a sum of Gaussian functions
with variance $\sigma^2$, centered on each of the 
neighbors of the central atom,
as well as on the central atom itself:
\begin{equation}\label{eq:atomic-density}
\rho_\mathcal{X}(\mathbf{r}) = \sum_{i\in\mathcal{X}} 
\exp\left(-\frac{(\mathbf{x}_i-\mathbf{r})^2}{2\sigma^2}\right).
\end{equation}
The SOAP kernel is then defined as the overlap of the two local atomic
neighbour densities, integrated over all three-dimensional rotations $\hat R$,
\begin{equation}
\tilde{k}(\mathcal{X},\mathcal{X}')=
\int \D\hat{R} \left| \int \rho_\mathcal{X}(\mathbf{r}) \rho_\mathcal{X'}(\hat{R}\mathbf{r}) \D \mathbf{r}\right|^n.
\label{eq:soap-kernel-nonorm}
\end{equation}
Note that in the $n=1$ case the two integrals can be switched, and therefore the kernel
looses all angular information, so we focus on the $n=2$ case exclusively. 
For most applications it is helpful to normalise the kernel 
so that the self-similarity of any environment is unity, giving the final kernel
\begin{equation}
k(\mathcal{X},\mathcal{X}')= \tilde{k}(\mathcal{X},\mathcal{X}')/
\sqrt{\tilde{k}(\mathcal{X},\mathcal{X})\tilde{k}(\mathcal{X}',\mathcal{X}')}
\label{eq:soap-kernel-int}
\end{equation}
It is a remarkable property of the SOAP kernel that the integration over all rotations can be carried out analytically. First, the atomic neighbour density is expanded in a basis composed of spherical harmonics and a set of orthogonal radial basis functions $\{g_b(r)\}$,
\begin{equation}
\rho_\mathcal{X}(\mathbf{r}) = \sum_{blm} c_{blm} g_b(|{\mathbf r}|) {\mathrm{Y}}_{lm}(\hat {\mathbf r}),
\end{equation}
then the rotationally invariant {\em power spectrum} is given by
\begin{equation}
p(\mathcal{X})_{b_1 b_2l} =\pi\sqrt{\frac{8}{2l +1}} \sum_{m} (c_{b_1lm})^\dag c_{b_2lm}.
\end{equation}
Collecting the elements of the power spectrum into a unit-length vector $\hat{\mathbf p}(\mathcal{X})$, the SOAP kernel is shown\cite{bart+13prb} to be given by
\begin{equation}
k(\mathcal{X},\mathcal{X}')=\hat{\mathbf p}(\mathcal{X})\cdot \hat{\mathbf p}(\mathcal{X'})
\end{equation}
eventually leaving a definition of the distance as 
\begin{equation}
d\left(\mathcal{X},\mathcal{X}'\right)=\sqrt{2-2\hat{\mathbf{p}}(\mathcal{X})\cdot \hat{\mathbf{p}}(\mathcal{X}')}
\label{eq:soap-distance}
\end{equation}
The  SOAP kernel can be written in the form of a dot product,
therefore it is manifestly positive definite, which implies that 
the distance function~\eqref{eq:soap-distance} is a proper metric. 

\subsection{From local descriptors to structure matching \label{sec:global}}

The vectors that enter the definition of the environments 
are defined in such a way that their dot product is the 
overlap of (smoothed) atomic distributions. Given two structures
with the same number $N$ of atoms, we can compute an
\emph{environment covariance matrix} 
that contains all the possible pairings of environments
\begin{equation}
C_{ij}(A,B)=k\left(\mathcal{X}^A_{i},\mathcal{X}^B_{j}\right),
\end{equation}
This matrix contains the complete information on the pair-wise 
similarity of all the environments between the two systems. 
Based on it, one can introduce a global kernel
to compare two structures or molecules. We will discuss
and compare four different approaches. All of them are
meant to be normalized, i.e. the given expressions 
for $K(A,B)$ are to be divided by $\sqrt{K(A,A) K(B,B)}$
whenever the kernel is not normalized to one by construction.

\paragraph*{Average structural kernel} 
A first possibility to compare two structures involves computing 
an \emph{average kernel}
\begin{equation}
\begin{split}
   \bar{K}(A,B) =& \frac{1}{N^2} \sum_{ij} C_{ij}(A,B) =\\
   =&
   \left[\frac{1}{N}\sum_{i}\mathbf{p}(\mathcal{X}^A_{i})\right] 
   \cdot 
   \left[\frac{1}{N}\sum_{j}\mathbf{p}(\mathcal{X}^B_{j})\right] .
\end{split}
\label{eq:k-avg}
\end{equation}
One sees that $\bar{K}$ can be computed inexpensively by just storing
the average SOAP fingerprint between all environments of the two structures.
This kernel is also positive-definite, being based on a scalar product~\cite{berg84book}, 
and therefore induces a metric $\bar{D}(A,B)=\sqrt{2-2\bar{K}(A,B)}$. 
On the other hand, it is not a very sensitive metric: 
 two very different structures can appear to be the same 
if they are composed of environments 
that give the same fingerprint upon averaging. 

\paragraph*{Best-match structural kernel}
Another possibility, that has been used previously with different kinds of structural
fingerprints~\cite{De2011,sade+13jcp,De2014,rupp+07jcim} is to identify the best match between the environments
of the two structures,
\begin{equation}
   \hat{K}(A,B) = \frac{1}{N} \max_{\pi} \sum_{i} C_{i\pi_i}(A,B)
   \label{eq:k-match}.
\end{equation}
which can be accomplished with an $\mathcal{O}(N^3)$
effort using the Munkres algorithm~\cite{kuhn55nrlq}.
The corresponding distance has the properties of a metric, 
which means it can still be safely used to assess similarity 
between structures and molecules.
Unfortunately, this ``best-match'' kernel is not 
guaranteed to be positive-definite, which makes it 
less than ideal for use in machine-learning
applications. Furthermore, the distance obtained 
by a best-match strategy is continuous, but has 
discontinuous derivatives whenever the matching 
of environments changes. 
These problems can be solved or alleviated
by matching the environments based on a different strategy,
that combines features of the average and the best-match 
kernels. 

\paragraph*{Regularized entropy match kernel}
The best match problem can be also stated in an alternative 
form, namely 
\begin{equation}
   \hat{K}(A,B) =  \max_{\mathbf{P}\in \mathcal{U}(N,N)} \sum_{ij} C_{ij}(A,B) P_{ij}
   \label{eq:k-match-b}.
\end{equation}
where $\mathcal{U}(N,N)$ is the set of $N\times N$
(scaled) doubly stochastic matrices,
whose rows and columns sum to $1/N$,
i.e. $\sum_i P_{ij} = \sum_j P_{ij}=1/N$. 
We can then borrow an idea that was recently 
introduced in the field of optimal 
transport\cite{cutu13nips} to \emph{regularize} this 
problem, adding a penalty that instead aims 
at maximizing the information entropy for the matrix
$\mathbf{P}$ subject to the aforementioned
constraints on its marginals. 
Such ``regularized-entropy match'' (REMatch) 
kernel is defined as 
\begin{equation}
\begin{split}
   &\hat{K}^{\gamma}(A,B) = \operatorname{Tr}\mathbf{P}^\gamma\mathbf{C}(A,B), \\
   &\mathbf{P}^\gamma =\operatorname*{argmin}_{\mathbf{P}\in \mathcal{U}(N,N)} \sum_{ij}   P_{ij} \left(1-C_{ij}+\gamma \ln P_{ij}\right),
   \end{split}
   \label{eq:k-regmatch}
\end{equation}
where the regularization is given by an entropy term $E(\mathbf{P})=-\sum_{ij}P_{ij}\ln P_{ij}$. 
$\mathbf{P}^\gamma$ can be computed very efficiently, 
with $\mathcal{O}(N^2)$ effort, 
by the Sinkhorn algorithm~\cite{cutu13nips}
(see Appendix~\ref{app:sinkhorn}).
For $\gamma\rightarrow 0$, the entropic
penalty becomes negligible, and 
$\hat{K}^{\gamma}(A,B)\rightarrow \hat{K}(A,B)$. 
For $\gamma\rightarrow \infty$,  
one selects the $\mathbf{P}$ with 
the least information content, that is one with constant 
$P_{ij}=1/N^2$. Hence, in this limit 
$\hat{K}^{\gamma}(A,B)\rightarrow \bar{K}(A,B)$.

\paragraph*{Permutation structural kernel}
For the sake of completeness, we also discuss a 
fourth option: rather than summing over all possible
pairs of environments, one can consider each pairing
of environments separately, and sum over all the 
$N!$ possible permutations that define the pairings.
In order to kill off more rapidly the combinations
of environments that contain bad matches, one can
\emph{multiply} the kernels that appear in each pairing,
and define a \emph{permutation kernel}
\begin{equation}
   \breve{K}(A,B) = \frac{1}{N!} \sum_{\pi}\prod_{i} C_{i\pi_i}(A,B) = \operatorname{perm}{\mathbf{C}(A,B)}. 
   \label{eq:k-perm}
\end{equation}
This choice corresponds to the evaluation of the permanent
of the environment kernel matrix, and has some appeal 
as it is guaranteed to yield a positive-definite kernel~\cite{cutu07proc}. 
The evaluation of the permanent of a matrix, however, has 
combinatorial computational complexity\footnote{Although stochastic algorithms do exist
to compute it to a desired precision in polynomial 
time~\cite{jerr+04jacm}}.
Its application is limited to small molecules, and we will
not discuss it further in the present work.

\subsection{Matching structures containing multiple species}

When comparing structures that contain different atomic
species, the first problem that has to be addressed is that
of extending the local environment metric so that the presence
of multiple elements is properly accounted for. 

SOAP descriptors provide a straightforward way to do this:
a separate density can be built for each atomic species
\begin{equation}
\rho_\mathcal{X}^\alpha(\mathbf{r}) = \sum_{i\in\mathcal{X}_\alpha} 
\exp\left(-\frac{(\mathbf{x}_{i}-\mathbf{r})^2}{2\sigma^2}\right),
\end{equation}
and a (non-normalized) kernel be defined by matching 
separately the different species:
\begin{equation}
\begin{split}
\tilde{k}(\mathcal{X},\mathcal{X}')=&
\int \D\hat{R} \left| \int \sum_\alpha \rho_\mathcal{X}^\alpha(\mathbf{r}) \rho_{\mathcal{X}'}^{\alpha}(\hat{R}\mathbf{r}) \D \mathbf{r}\right|^2\\
=&\sum_{\alpha\beta} \mathbf{p}_{\alpha\beta}(\mathcal{X})\cdot
\mathbf{p}_{\alpha\beta}(\mathcal{X}'). \label{eq:k-multispecies}
\end{split}
\end{equation}
Here we have introduced ``partial'' power spectra $\mathbf{p}_{\alpha\beta}$ that encode
information on the relative arrangement of pairs of species, and can be
written as 
\begin{equation}
p(\mathcal{X})^{\alpha\beta}_{b_1 b_2l} = \pi\sqrt{\frac{8}{2l +1}}\sum_{m} (c^{\alpha}_{b_1lm})^\dag c^\beta_{b_2lm},
\label{eq:pwrMulti}
\end{equation}
where we built in the angular channel dependent weights into the elements of the power spectrum. The expansion coefficients describe the atomic density of species $\alpha$ 
\begin{equation}
\label{eq:atomicDensitySpecies}
\rho_\mathcal{X}^\alpha(\mathbf{r}) = \sum_{blm} c^\alpha_{blm} g_b(|{\mathbf r}|) {\mathrm{Y}}_{lm}(\hat {\mathbf r})
\end{equation}
in terms of a basis set, which is a combination of spherical harmonics and orthogonal radial functions. 
 The kernel in Eq.~\eqref{eq:k-multispecies} can then be
normalized as in Eq.~\eqref{eq:soap-kernel-int}. 

Note that, even though the overlap between the
environments of the different species is considered 
to be zero, the kernel is sensitive the relative 
correlations of different species. This is because,
due to the squaring of the density 
overlap within the rotational average,
the SO(3) power spectrum vectors 
contain mixed-species components. 
One could also introduce a notion of ``alchemical 
similarity'' between different species. For instance, when
comparing structures of III-V semiconductors one could 
disregard the chemical information on the identity of 
an atom as long as it belongs to the same column of the 
periodic table. 
Such a notion can be readily implemented, 
defining an alchemical similarity kernel $\kappa_{\alpha\beta}$ 
which is one for pairs that should be considered 
interchangeable, and tend to zero for pairs that 
one wants to consider as completely unrelated. 
The expression then becomes
\begin{equation}
\label{eq:alchemicalKernel}
\begin{split}
\tilde{k}(\mathcal{X},\mathcal{X}')=&
\int \D\hat{R} \left| \int \sum_{\alpha\alpha'} \kappa_{\alpha\alpha'} \rho_\mathcal{X}^\alpha(\mathbf{r}) \rho_{\mathcal{X}'}^{\alpha'}(\hat{R}\mathbf{r}) \D \mathbf{r}\right|^2\\
=&
\sum_{\alpha\beta\alpha'\beta'} \mathbf{p}_{\alpha\beta}(\mathcal{X})\cdot
\mathbf{p}_{\alpha'\beta'}(\mathcal{X}')\kappa_{\alpha\alpha'} \kappa_{\beta\beta'}.
\end{split}
\end{equation}
The original expression~\eqref{eq:k-multispecies}
can be recovered by setting $\kappa_{\alpha\beta}=\delta_{\alpha\beta}$. 
Global similarity kernels can then be transparently introduced to 
compare structures composed of different atomic species, with geometry
and alchemical composition treated on the same footings and the possibility
of adapting the definition of similarity to the system and application.

\subsection{Matching structures with different numbers of atoms}

The definitions above can be readily extended to compare structures
containing different numbers of atoms $N_A$ and $N_B$. We discuss two possible
strategies. When comparing crystalline, periodic structures, it may be the
case that one of the structures corresponds to a slight distortion of the 
other, that needs a larger unit cell for a proper representation.
Comparing the structures using the average kernel~\eqref{eq:k-avg}
does automatically the ``right thing'', that is performing the 
comparison in a way that is independent of the number of times 
the two structures have to be replicated to match atom counts.
 In the case of the permutation kernel and of the best-match
 kernel, the most effective way to perform the 
 comparison is to evaluate the least common multiple $N$ of $N_A$ and $N_B$, and
replicate the environment similarity matrix to form a square
matrix. One can then proceed to compute the permanent,
or the linear assignment problem, based on such replicated
matrix. The advantage of this procedure is that one does not
need to explicitly
find the relation between the shape of the two unit cells and replicate
them to perform the comparison: the environment similarities can be
evaluated including periodic replicas, and the minimum number of 
comparisons will be naturally performed among any pairs of structures. 
However, the least common multiple can become very large, 
making even the best-match kernel~\eqref{eq:k-match} 
impractically demanding, although the cost can be reduced by exploiting the
redundancy in the extended environment covariance matrix.
As shown in the Appendix, the REMatch kernel~\eqref{eq:k-regmatch}
can be computed easily also for a rectangular matrix,
which constitutes an additional advantage of formulating
the environment matching problem in terms of a regularized 
transport optimization.

When comparing molecules or molecular fragments, it may
be advisable to proceed differently -- since in that case
the chemical composition might differ, and it may not make
sense to compare molecules as if they were part of an infinite
periodic assembly. 
A possible strategy is then to consider, given a molecular
database, the smallest pool (``kit'') of atoms from which 
every molecule in the set can be constructed. Then,
when comparing each pair of structures, the atoms that are
not needed to form either of the two molecules would still
be part of the comparison, in the form of idealized 
``isolated'' species. Alternatively, for instance
when the full database is not known a priori, 
such ``reference kit'' could be chosen dynamically for 
each pair of molecules.
Since the SO(3) fingerprints that underlie the definition of 
the SOAP kernel can also be evaluated for isolated atoms\footnote{The density of atoms defined in equation \eqref{eq:atomic-density} contains the central atom.},
it is then possible to introduce a natural definition 
of the covariance between an environment and an isolated
atom. One of the advantages of such approach is that the 
global kernels will then vary smoothly if a molecule is
continuously broken up into its constituent atoms, which
lends itself to a very effective description of atomization
processes.

\begin{figure*}
\centering
\includegraphics[width=1.0\textwidth]{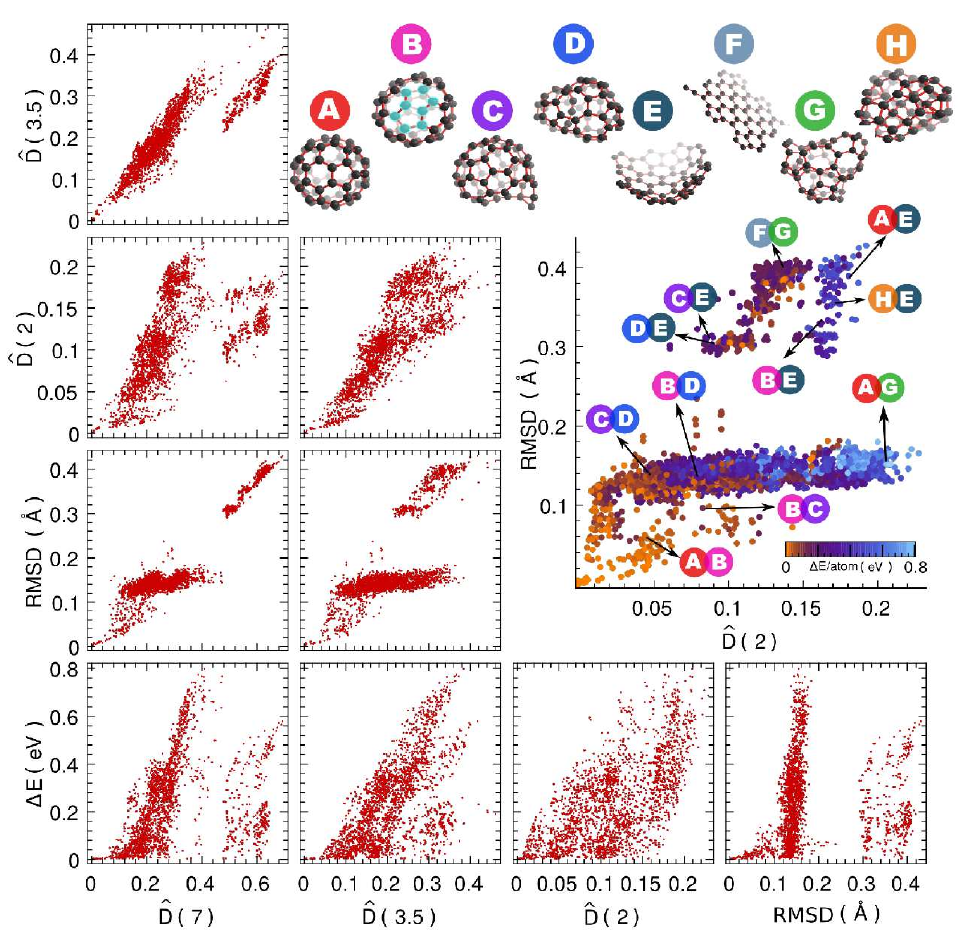}
\caption{
The figure compares the value of global structural similarities
for different pairs of structures taken from the 
80 local energy minima of C$_{60}$ discussed in Ref.~\cite{De2011}
The structural similarities considered include
the absolute difference in energy per atom, the 
(permutation invariant\cite{sade+13jcp}) RMSD per atom, 
and the best-match combination of SOAP kernels 
computed with different cutoff distances
(2\AA{}, 3.5\AA{}, 7\AA{}). 
The correlation between RMSD and $\hat{D}$ based
on 2\AA{}-cutoff SOAP is enlarged, color-coded based on
energy differences and annotated with selected pairs of 
structures corresponding to different distances.
} 
\label{fig:C60-CORR}
\end{figure*}

\subsection{Representing (al)chemical landscapes}

In this work we will demonstrate the flexibility, transferability
and effectiveness of the framework we have just introduced to compare
molecular and condensed-phase structures. To this aim, we will build 
two dimensional maps that represent proximity relations between the 
structures -- as assessed by the kernel-induced metric -- using 
sketch-map~\cite{ceri+11pnas}, a non-linear
dimensionality reduction (NLDR) scheme specifically designed 
to deal with atomistic simulation data. 
As we will demonstrate, the combination
of SOAP-based structural metrics and NLDR representation provides a
broadly applicable protocol to generate an insightful representation 
of the structural and alchemical landscape of complex molecular and
condensed-phase systems. Of course, one could use
the SOAP-based global kernels, or the corresponding distances,
as the basis of other non-linear dimensionality reduction
techniques, such as multi-dimensional scaling~\cite{cox-cox10boox}
or diffusion maps~\cite{coif+05pnas,ferg+10pnas,rohr+arpc13}.

We refer the reader to the relevant literature for a
detailed explanation of the sketch-map
algorithm\cite{ceri+11pnas,trib+12pnas,ceri+13jctc}. The main idea
derives from multi-dimensional scaling, and is based on optimizing
a non-linear objective function
\begin{equation}
S^2 = \sum_{ij} \left[F\left[D(X_i,X_j)\right]-
f\left[d(x_i,x_j)\right]\right]^2 \label{eq:smap}
\end{equation}
where $\left\{X_i\right\}$ and $\left\{x_i\right\}$ correspond 
respectively to high-dimensional reference structures
and to vectors in a low-dimensional space. The metric $d$ in 
low dimension is typically taken to be the Euclidean distance,
whereas the metric in high dimension could be more complex. In
this case, $X_i$ can be regarded as an abstract descriptor of 
a structure or molecule, and $D$ is one of the kernel-based distance 
metrics discussed above.
$F$ and $f$ are non-linear sigmoid functions of the form
\begin{equation}
F(r) = 1 - ( 1 + (2^{a/b} - 1)(r/\sigma)^a )^{-b/a},
\end{equation}
which serve to focus the optimization of~\eqref{eq:smap} on
the most significant, intermediate distances, disregarding local
distortion (e.g. induced by thermal fluctuations) and the relation
between completely unrelated portions of configurational landscape.
The choice of the parameters in the sigmoid functions 
is discussed in Ref.~\cite{ceri+13jctc}. Here we will label
synthetically each sketch-map representation using the
notation {\texttt{$\sigma$-A\_B-a\_b}} where $A$ and $B$ denote
the exponents used for the high-dimensional function $F$,
$a$ and $b$ denote the exponents for the low-dimensional function $f$,
and $\sigma$ the threshold for the switching function.
Open-source software to perform the dimensionality-reduction
step, as well as to compute the different similarity kernels
we have introduced, is available from the authors upon request. 
Interactive versions of the structural maps discussed
in the text are provided in the supporting information (SI),
and are available as an on-line repository~\cite{sketchmaporg}.

\begin{figure*}[tbhp]
\centering
\includegraphics[width=1.0\textwidth]{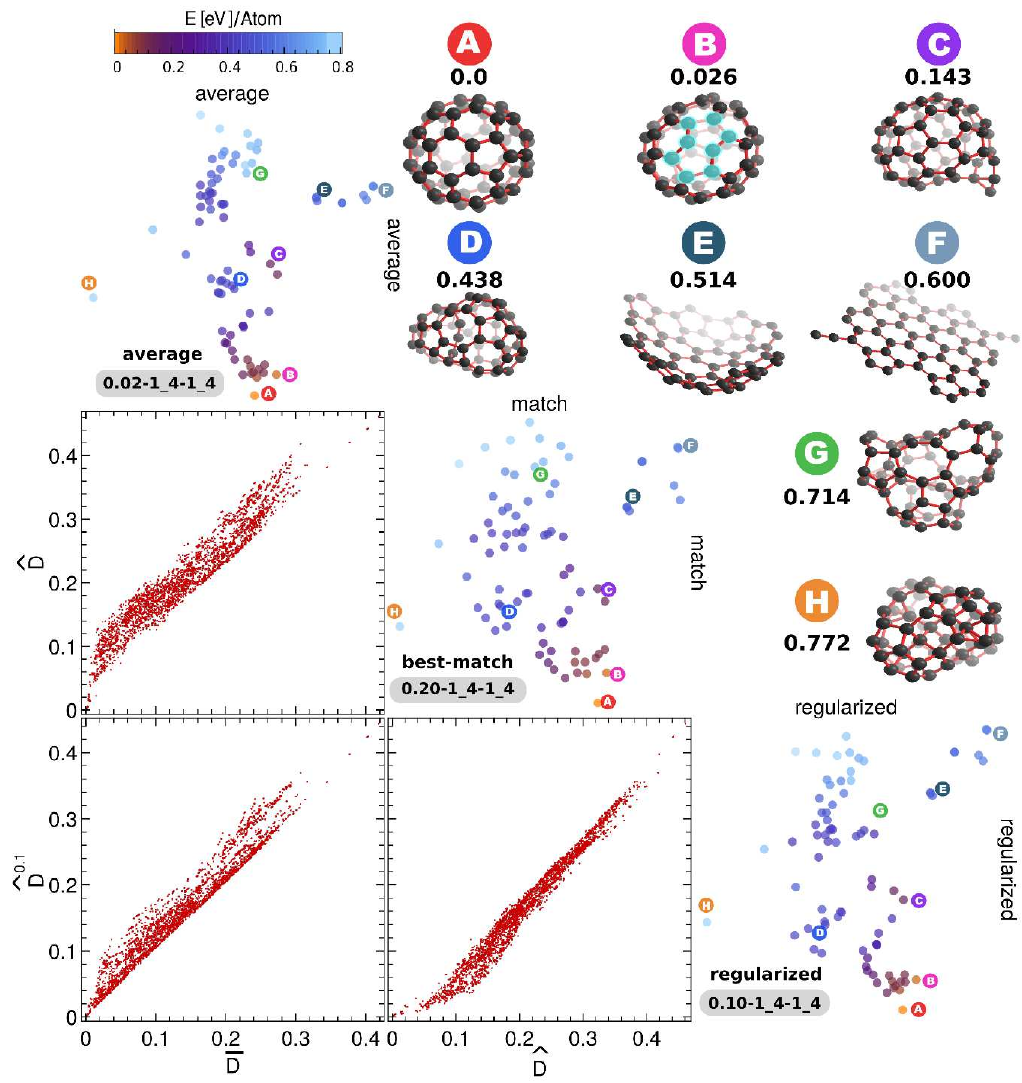}
\caption{
The figure compares the value of global structural distances
induced by the average, best-match, and REMatch kernels discussed in Section~\ref{sec:global}, for 
80 local-minimum structures of C$_{60}$. On the diagonal
we report the sketch-map projections of the structural landscape
based on the three metrics, colored according to the energy
of each structure, as obtained by Sandip et al. ~\cite{De2011}.
Eight representative structures and their positions on the Sketch-maps have been indicated with letters on color coded disks. The numeric value on the top of each structure represents their energy in eV, relative to the global minimum. SOAP descriptors were computed using a cutoff of 3.5\AA{} and the Sketch-map parameters are indicated on the map according to the syntax described in the text.
} 
\label{fig:C60}
\end{figure*}

\section{Examples and applications}

After having described the theoretical and algorithmic
background of our strategy to define a structural similarity
kernel, let us present a series of applications.
In order to demonstrate that our approach can be seamlessly
applied to the most diverse atomistic simulation problems, we have chosen
examples of increasing complexity, from clusters, to 
crystalline and amorphous solids, to biological molecules
and a database of small organic compounds, containing varying number of both atoms and atomic species. 
As we will discuss, SOAP-based structural kernels contain 
several adjustable hyperparameters, that can be regulated to focus 
the  dissimilarity measure onto the desired features. Unless
otherwise specified, however, we have not explored fully this 
possibility, and we have simply chosen reasonable values of the
parameters without much fine-tuning.

\subsection{The energy landscape of C$_{60}$ clusters}

Let us start with a relatively simple test case. 
We consider the same set of 80 local minima for C$_{60}$ 
discussed in Ref.~\cite{De2011}, 
which were obtained by exploring the Density Functional Theory energy 
landscape of C$_{60}$ using the Minima Hopping ~\cite{minhop} global structure 
search algorithm.  Figure:~\ref{fig:C60-CORR} contrasts different
similarity matrices: the permutation-invariant RMSD\cite{sade+13jcp},
the absolute difference between the potential energy, and
the best-match distances obtained from SOAP descriptors 
computed with different environment cutoff. 
RMSD distance does not correlate very well with 
SOAP-based metrics, particularly for the smaller cutoff
value. The $\bar{D}(2\text{\AA})$-RMSD correlation plot is 
enlarged, and allows us to discuss the source of this discrepancy. 
Hollow fullerene-like structures 
(A, with reference to the labeling
in the figure) and compact structures 
containing internal connections
(A,G) are extremely different from the point 
of view of the short-range
connectivity, but  differ comparatively less in terms of RMSD,
since they are both fairly compact. On the other hand, flake-like
structures based on a honeycomb motif (F,E) have the same basic
first-neighbor connectivity as the defective fullerene structures (C,D) 
but have much different spatial extent.
Then, one sees that the discrepancy between RMSD and small-cutoff
$\hat{D}$ indicates just the focus on different structural features:
the global arrangement of atoms in the first case, and the local 
connectivity in the latter. In the case of SOAP-based metrics, however
it is easy to extend the sensitivity of the metric to longer distances
just by increasing the cutoff: by going from 2\AA{} to 3.5 and 7, 
one sees that $\hat{D}$ and RMSD become progressively more 
correlated, as the focus shifts from the nearest-neighbor 
coordination to the overall geometry of the cluster.

 \begin{figure*}[tbhp]
\centering
\includegraphics[width=\textwidth]{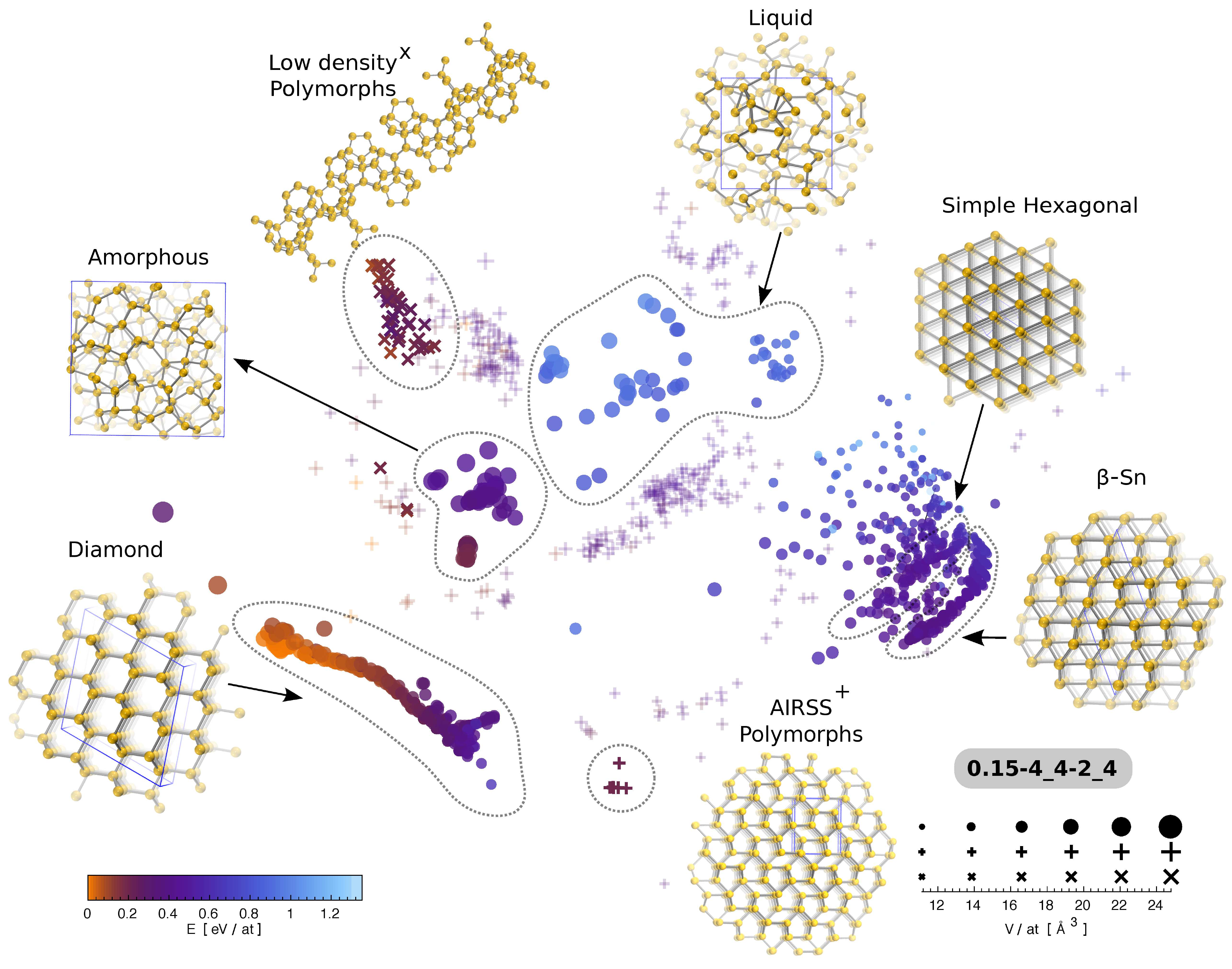}
\caption{Sketch-map of 1274 crystalline and amorphous silicon structures 
obtained by sampling different phases from the
phase diagram (disks), polymorphs obtained by ab initio random 
structure search\cite{airss} ($+$ signs) and by minima hopping\cite{sipoly}  ($\times$ signs). 
The color and size of the points varies according to their 
atomic energy and atomic volumes respectively. Regions 
of the plot which represents different phases have been outlined 
with dotted contours. 
\label{fig:silicon}
}
\end{figure*}

It is worth stressing that the RMSD, albeit a very natural 
measure of structural similarity, is not necessarily the best 
metric to compare configurations. 
To see why, consider the absolute energy difference as 
a measure of similarity: even though one can obviously 
have configurations with very different geometry and similar 
energies, in general one would expect that on the contrary large 
energy differences should be
associated with highly dissimilar structures in a given system 
-- which is not the case for RMSD. 
One sees that the intermediate-cutoff $\bar{D}(3.5\text{\AA})$
shows a nice correlation between energetic and structural differences.
   
These considerations underline a theme that will recur 
in other examples: SOAP-based structural metrics offer a 
mathematically sound framework 
that can be transparently adapted to focus on the aspects that are
most relevant to a given application. For instance, power-spectrum
based environment kernels are invariant to mirror symmetry, and 
therefore the derived metrics cannot distinguish enantiomers. 
If one needed to do so, however, it would be sufficient to use
a bispectrum-based SOAP kernel~\cite{bart+13prb} -- 
which corresponds to $n=3$ in 
eq. (\ref{eq:soap-kernel-nonorm}) and is 
invariant to rotations but not to mirror symmetry operations --
as the basis for obtaining a global comparison
that is sensitive to chirality.
   
Having established a connection between traditional structural
similarity metrics and the best-match SOAP kernel, let us use 
the example of C$_{60}$ to compare the three main strategies 
we propose to build a global kernel: the average kernel $\bar{K}$, 
the best-match kernel $\hat{K}$, and the regularized entropy match kernel  
$\hat{K}^\gamma$ with an intermediate regularization parameter
$\gamma=0.1$.

The distance-distance correlation plot for each pair of structures, 
that compares the distances induced by the three kernels, 
is reported in Fig.\ref{fig:C60}. 
The $\bar{D}-\hat{D}$ plot shows overall linear correlation 
except for very small values of $\bar{D}$. 
This is expected as the average kernel is under-determined, and
could in principle label two structures as identical even though 
they might be composed of different environments. The best-match 
kernel, therefore, provides better resolving power. 
As we will discuss in more detail later on, the regularized
best-match kernel $\hat{D}^\gamma$ can be 
tuned to interpolate between these
two extremes. As an example, we chose here an intermediate
value $\gamma=0.1$: as shown in Fig. \ref{fig:C60}, the
resulting distance correlates strongly with both $\bar{D}$
and the conventional best-match distance $\hat{D}$.

Fig.\ref{fig:C60} also shows annotated sketch-maps obtained
based on the three metrics. 
Once the sketch-map parameters have 
been adjusted following the guidelines in Ref.~\cite{ceri+13jctc}, 
the three maps are effectively equivalent -- indicating
that the three kernels give similar qualitative information on the similarity 
between different structures. Given the much lower
computational cost associated with the evaluation of the average
kernel, this observation suggests it might conveniently be used to 
preliminarily screen a dataset before proceeding to a more accurate
comparison of similar structures based on the best-match, or REMatch distance.

\subsection{Natural and hypothetical polymorphs
of silicon}

As a second example, let us move on to a condensed-phase 
application. Here we start from a database of 1274 bulk silicon 
structures containing ideal and distorted configurations 
from the phase diagram (e.g. diamond, simple hexagonal, 
$\beta$-tin, liquid and quenched amorphous structures).
SOAP environment kernels with a 5 \AA{} cutoff distance 
were used, and combined with a best-match strategy 
to obtain the (dis)similarity matrix 
\footnote{Some of the structures in the overall data set have
numbers of atoms in the unit cell that would lead to a 
large least common multiple when repeating the environment
similarity matrix to form a square matrix. One can keep
the cost of computing the similarity matrix low by 
exploiting the redundancy in the similarity matrix,
or by approximating $\hat{K}$ using a REMatch kernel 
with a small entropy regularization.}
We selected 100 landmark configurations out of this 
data set (using farthest point sampling based on 
kernel distance) and built a sketch-map, on which
the rest of configurations were projected.
The outcome of such mapping procedure is shown in Fig.~\ref{fig:silicon},
where points are colored according to the DFT atomic energy,
and point sizes are scaled to a size proportional to volume per atom. 
As seen in the Fig.~\ref{fig:silicon} the map is extremely well 
correlated with both atomic energy and density. 
Furthermore, structures that were obtained by distorting and 
heating up structures coming from different portions of the 
phase diagram are clustered together: rough outlines have 
been drawn on the map to indicate different phases.

Although the map has been built using only reference 
configurations from a few of the conventional Si phases, we have also projected
on it (using out-of-sample embedding) two sets of 
hypothetical configurations obtained by minima hopping~\cite{sipoly} and by ab initio random structure search (AIRSS)~\cite{airss,rapp-ncom-2015}. These structures were not included
in the landmarks selection phase. Still, the out-of-sample
embedding procedure correctly identifies not only that in 
most cases AIRSS structures differ significantly from 
stable phases of silicon, but also clusters together
hypothetical polymorphs that share common features.
For instance, the AIRSS structures outlined in the lower
portion of the map are all taken from Ref.~\cite{rapp-ncom-2015}. 
The structures were proposed as possible metastable polymorphs 
arising as a result of a microexplosion (induced by powerful, ultrashort
and tightly focused laser pulses) in crystalline cubic diamond 
silicon phase, hence their structural motif naturally 
carries resemblance with silicon diamond phase. It is 
interesting to see that they indeed are projected 
close to the diamond phase on the map.   
All of the minima hopping low-density Si polymorphs 
are also clustered together, which is consistent with 
the fact that they are all based on combinations of a few
base motifs. 
Thus, Figure~\ref{fig:silicon} shows not only that 
SOAP-based structural similarity distances can be very
effective in the study of bulk crystalline structures,
but also testifies the extrapolative power of a sketch-map
representation based on such a metric. 

\begin{figure*}[btph]
\centering
\includegraphics[width=\textwidth]{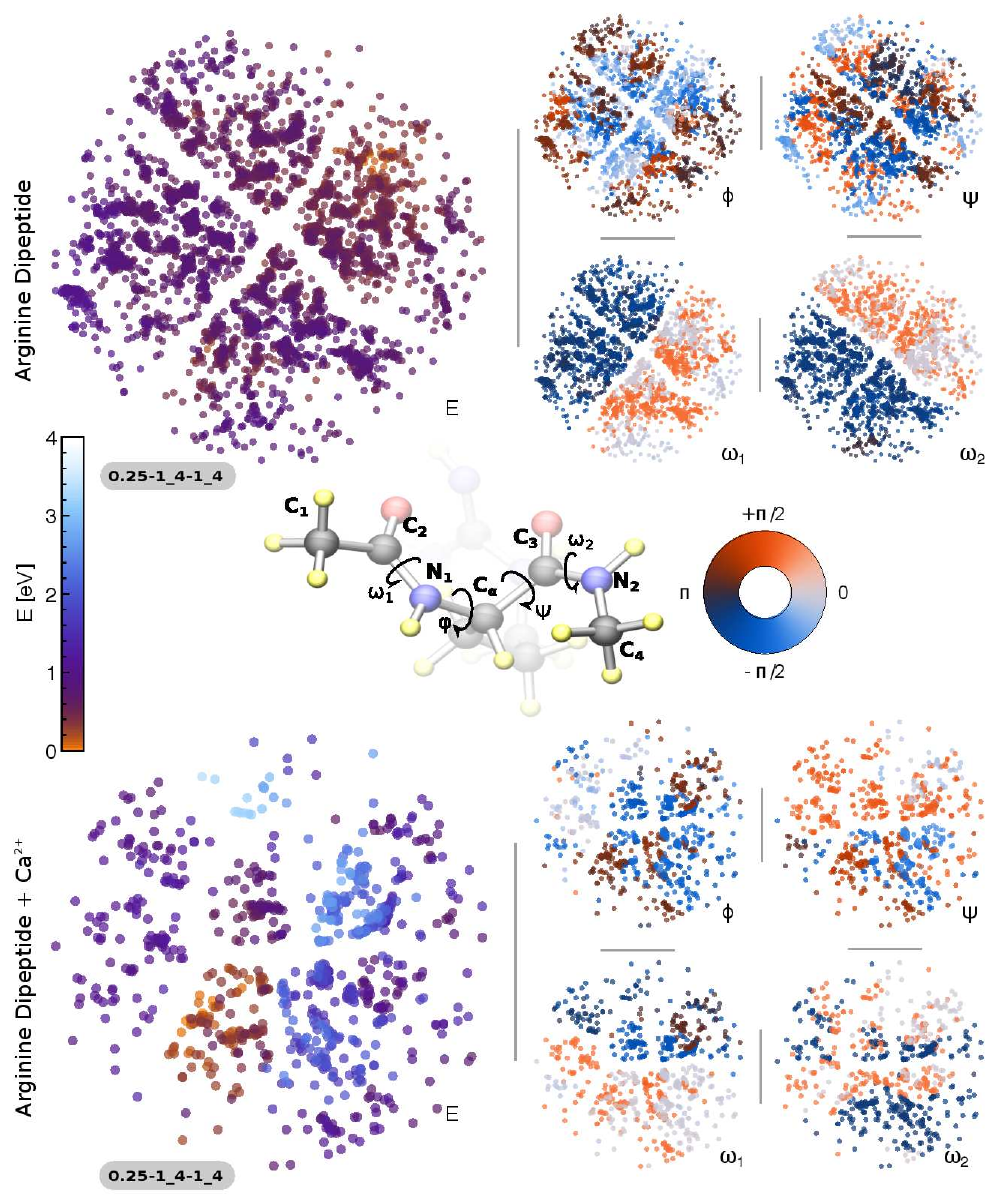}
\caption{Sketch-map representation of locally stable 
arginine dipeptide conformers, without (top) and with (bottom) a Ca$^{2+}$ ion.
Left-hand panels are colored according to the energy relative to the minimum energy form, while the smaller maps on the right are colored according to the values of different dihedral
angles, as indicated in the legend.
}
\label{fig:arg-cfg}
\end{figure*}

\subsection{Arginine Dipeptide}

Having shown that SOAP-based structural similarity kernels
are equally effective for clusters and for bulk configurations
of elemental materials, let us consider a case of a multi-species
chemical compound. We selected a library of 5062 locally stable 
conformers of arginine dipeptide (845 with and 4217 without a Ca$^{2+}$ 
counterion) from a public database of 
oligopeptides structures developed
by Ropo et al \cite{ropo+16sd}. 
We used a cut-off of 3.5\AA{} in the definition of environment
SOAP kernels, and combined them using a best-match strategy.
Since H atoms stay at almost fixed positions relative to their 
neighboring atoms, we decided to include them in the 
environment descriptors of other atoms, but did not include them
explicitly as centers of atomic environments. This is another
example of how SOAP-based structural metrics are effective in
a broad variety of contexts, but at the same time can be 
easily and transparently refined based on intuition, 
prior experience, or a clear understanding of the objectives
of the structural comparison.

In Fig.~\ref{fig:arg-cfg} we show the sketch-map representation
for these two sets of structures, highlighting the correlation
between the location on the map and structural and energetic
properties of the conformers. In the absence of a complexing 
cation, the dipeptide can exist in a very large number of 
local minima, spanning a relatively narrow range of energies. 
The map shows very clearly partitioning of configuration space
in four disconnected regions. 
Conventional wisdom~\cite{ramachandran-plot} assumes
that the C$_\alpha$ dihedral angles $\phi$ and $\psi$
are the most important descriptors of oligopeptide
structure. One quickly realizes, however, that
the order parameters corresponding to the
four lobes are connected to the cis-trans isomerization
of the two peptide bonds. Within each of the lobes, configurations
with different $\phi$-$\psi$ dihedral angles are clearly clustered
together, but in this case they constitute features of 
secondary importance. This observation demonstrates
the advantages of using a general-purpose descriptor,
that does not rely on pre-conceived assumptions on
the behavior of the molecule being studied, 
but instead captures automatically 
the intrinsic structural hierarchy of minima in the configuration landscape. 

The presence of a Ca$^{2+}$ cation has a dramatic
impact on the landscape for the dipeptide. The distribution of 
configurations becomes considerably more sparse and spans a broader
range of energies. The strong electrostatic interaction with the 
cation means that there is not a clear separation anymore between
the energy scale for $\phi$-$\psi$ flexibility of the backbone and
the isomerization of the peptide bonds. 

\begin{figure}[tbph]
\centering
\includegraphics[width=\columnwidth]{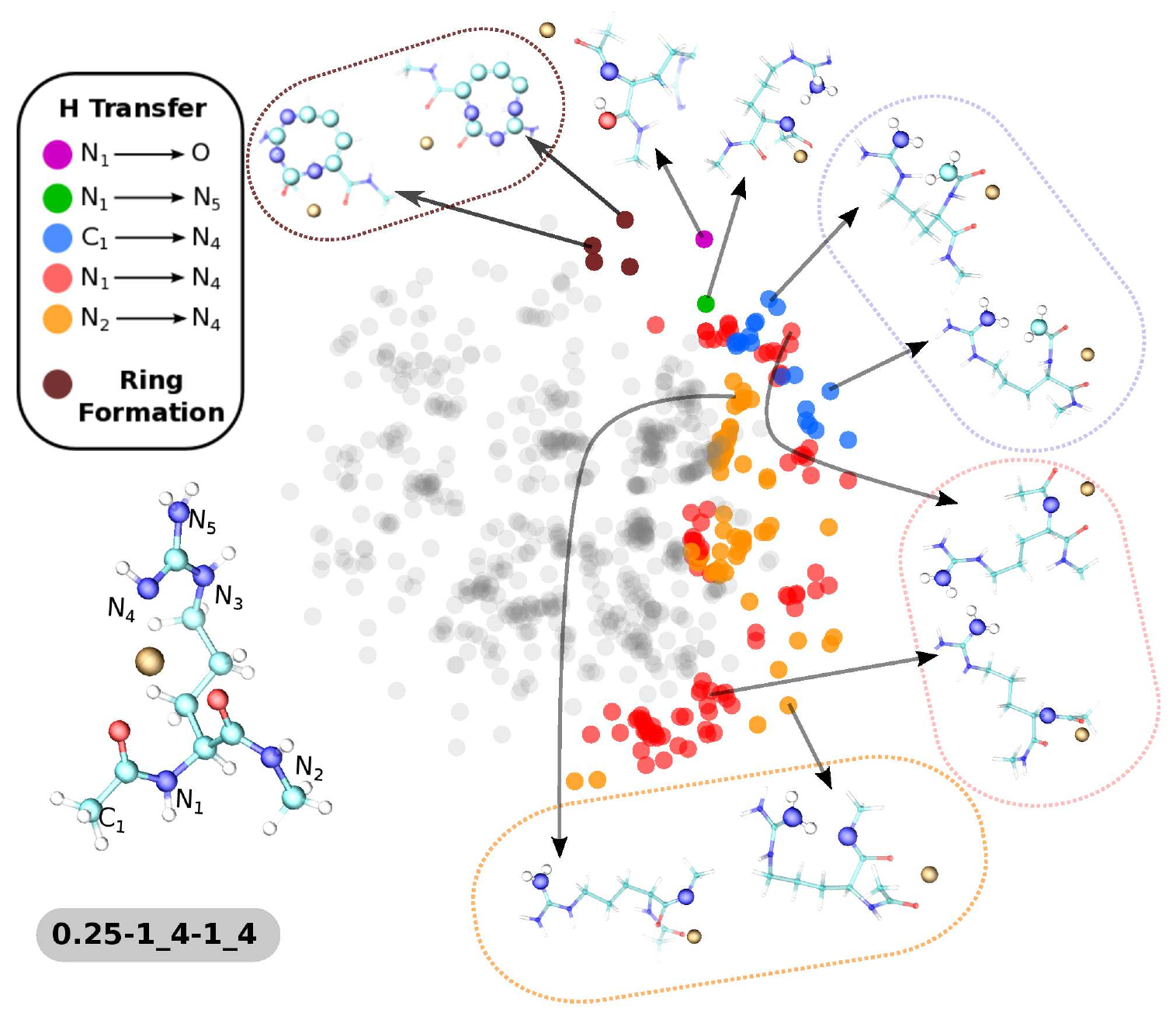}
\caption{Sketch-map representation of stable configurations of Arginine dipeptide complexed with a Ca$^{2+}$ ion. The 
structures that have undergone a proton transfer reaction
relative to the neutral molecule have been highlighted, and
a few representative snapshots of the molecular structure
are also reported.}
\label{fig:arg-chem}
\end{figure}

A remarkable observation in this analysis is the
realization that the presence of the cation catalyzed 
unexpected proton transfer reactions, that change the 
chemical structure of the molecule. Configurations that
underwent a chemical reaction are clustered on one side of the map
(Fig.~\ref{fig:arg-chem}),
with further internal structure reflecting the fact that 
SOAP-based structural metrics treat on the same footing 
information on the chemical bonding and on the conformational
variability of the molecule. It is again worth noting that 
by changing the cut-off value for the SOAP descriptors, one
can ``focus'' the structural metric on different molecular 
features. A short cutoff of 2\AA{} makes the chemically
different structures stand out more as outliers -- which would
for instance be useful to detect automatically this kind
of unwanted transitions in an automatically-generated data set --
while on the contrary a longer cutoff would give more 
importance to the difference between collapsed and extended
molecular conformers.

\begin{figure}[tbph]
\centering
\includegraphics[width=0.9\columnwidth]{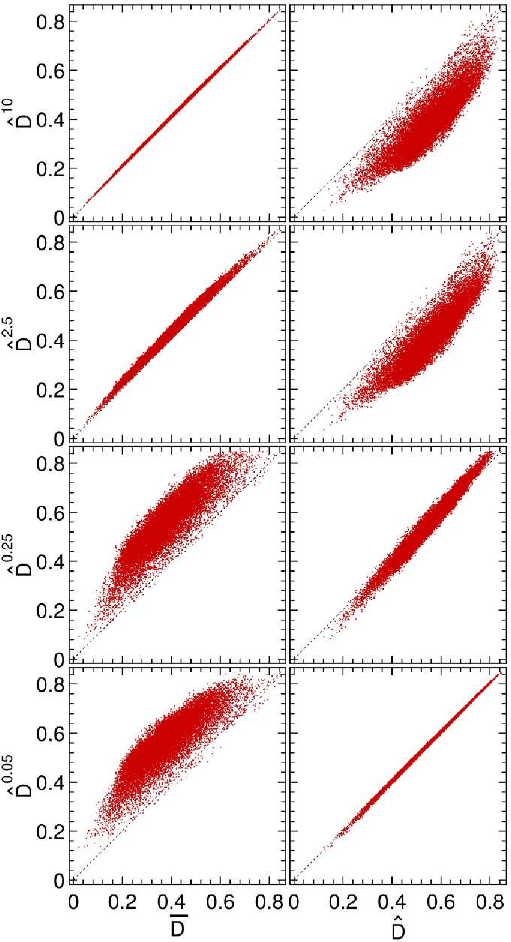}
\caption{Correlations between structural similarity
distances induced by the average kernel $\bar{K}$, 
the best-match kernel $\hat{K}$, and \emph{regularized}
best-match kernels $\hat{K}^\gamma$ with different
regularization parameters $\gamma$. Distances are 
computed between pairs of 200 structures, randomly
selected from the QM7b database\cite{qm7b,qm7b-father}.}
\label{fig:qm7b-corr}
\end{figure}

\begin{figure*}[tbph]
\centering
\includegraphics[width=0.9\textwidth]{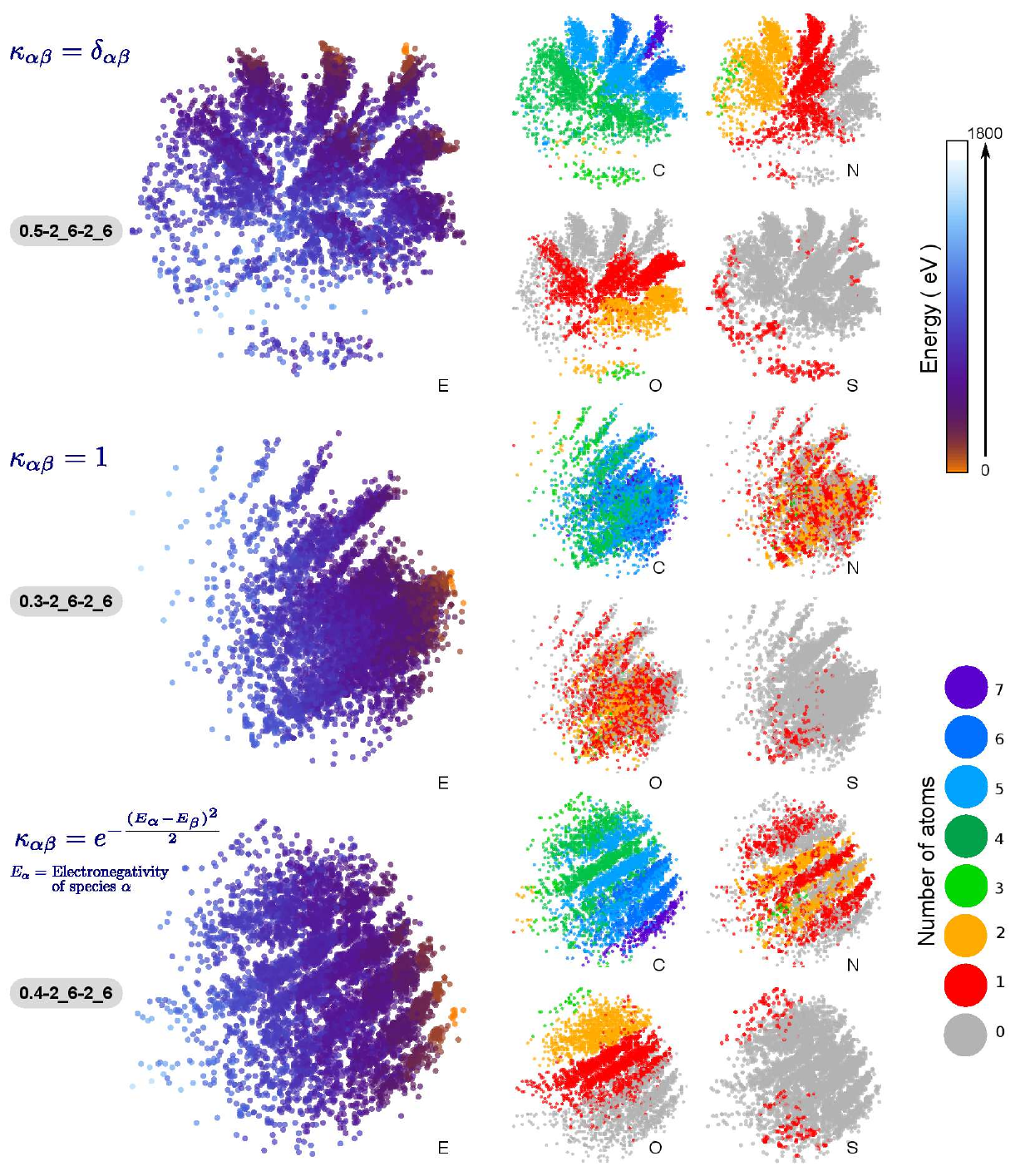}
\caption{Sketch-map representation of minimum-energy
structures from a database of molecules containing
up to 7 non-hydrogen atoms (C O N S Cl), and saturated with
hydrogen to varying degrees\cite{qm7b}. 
Left-hand panels
show the map colored according to the atomization
energy as computed by DFT. In the right-hand images,
the points are colored according to the number
of constituent C, O, N, S atoms. 
The top row corresponds to an alchemical kernel
that treats all species as different, the 
middle row treats all the non-H atoms as the 
same species, whereas the bottom row introduces
an alchemical kernel that depends on the 
difference in electronegativity between
species.}
\label{fig:qm7b}
\end{figure*}

\subsection{Mapping (al)chemical space}

As a final example of the evaluation of a 
structural and alchemical similarity metric, and its
use to represent complex ensembles of compounds, 
let us consider the QM7b database~\cite{qm7b}.
This set of compounds contains 7211 minimum-energy
structures for small organic compounds containing
up to seven non-hydrogen atoms (C, N, O, S, Cl), 
saturated with H to different degrees. This 
database constitutes a small fraction of 
a larger chemical library that contains millions
of hypothetical structures screened for accessible 
synthetic pathways~\cite{qm7b-father}. 

This is an extremely challenging data set to 
benchmark a structural similarity metric: molecules
differ by number of atoms, chemical composition, 
bonding and conformation. To simplify the description,
we decided to use 
SOAP descriptors with a cutoff of 3\AA{}, 
and to include H atoms in the environments but 
not as environment centers, to simplify the 
description -- considering also that in the case of arginine
dipeptide this choice did not prevent clear
identification of isomers that only differed by 
a proton transfer reaction. We used a best-match 
strategy to compare configurations, and topped 
them up with isolated atoms up to the maximum number
of each species that is present in the database. 
This effectively corresponds to choosing a 
``kit'' (in other terms, a fully atomized reference state)
starting from which all of the compounds can be assembled.

This is a fairly extreme case for the application of
our idea of compounding local structure matching to
obtain a global structural metric, so it is worth 
returning on a comparison of the different strategies
we proposed. Fig.~\ref{fig:qm7b-corr} compares
average and best-match distances to
REMatch kernels using different regularization
parameters. Despite the very different context, the
outcome is similar to what we observed in 
Figure~\ref{fig:C60} for C$_{60}$ clusters. The 
average kernel is reasonably well correlated
with the more demanding best-match kernel, although
in most cases it has poorer resolution. By 
varying $\gamma$, the regularized match distance
$\hat{D}^\gamma$ varies between these two extremes,
and for $\gamma<1$ provides a smooth, inexpensive
approximation to the best-match distance. 

For the sake of simplicity (and given we reduced the
size of the environment covariance matrix $\mathbf{C}$ 
not considering H atoms as environment centers) we used the 
conventional best-match distance for the rest of 
our analyses. 
As shown in Fig.~\ref{fig:qm7b}, the SOAP-based 
metric nicely separates out ``islands'' with homogeneous
composition in terms of the number of non-H atoms. 
Within each group of atoms, one can recognize some
sub-structure, with configurations roughly arranged in terms
of the atomization energy -- which in turns strongly
correlates with the degree of H saturation. 
As it can be seen from inspection of the database
(see the SI) in many cases one can notice that structures
with similar chemical skeleton (presence of cycles,
chemical groups, etc.) are clustered close to 
each other in the map. However, it is of course very
difficult to quantitatively assess how well the map
corresponds to chemical intuition, and how much 
departures from it are to be considered a failure of 
the metric, of the sketch-map procedure or of 
the notion of ``chemical intuition''. 

Our objective here is more to demonstrate how the
fingerprint-based structural metric we introduced
can cope with widely different classes of problems,
and how it can treat on the same footings alchemical and
structural variability. As an example 
we have also computed the similarity matrix
and mapped the QM7b landscape using a modified
alchemical similarity metric between the non-H
atoms (we always take $\kappa_{\alpha\text{H}}=\delta_{\alpha\text{H}}$). 
First, we set 
$\kappa_{\alpha\beta}=1$ (which means we are treating 
species $\alpha$ and $\beta$ as the same species) for all of  
atoms except H.
The clear separation of the map into islands with the
same stoichiometry is lost. However, there is now
near-perfect correlation between position on the map and 
atomization energy, and at the same time one can see some
residual clustering of molecules with similar composition. 
This can be explained because information on the 
alchemical identity of the atoms is encoded in their 
atomic coordination and bond lengths.
This is for instance evident for sulfur, 
that has considerably larger bond lengths, leading to 
better clustering of sulfur-containing compounds 
than in the case of oxygen or nitrogen. 

Obviously, assuming that all atom kinds
are interchangeable is an extreme choice, and it is 
hard to imagine circumstances in which this ``element 
agnostic'' metric would be advantageous over one that
exploited knowledge of the chemical identity of atoms. 
On the other hand, one could foresee to encode
information on the ``alchemical similarity'' using 
one of the many quantities chemists have used historically
to rationalize trends in reactivity across the periodic
table. As an example, we used the electronegativity $E_\alpha$ to define 
\begin{equation}
\kappa_{\alpha\beta}=e^{-(E_\alpha -E_\beta)^2/2\Delta^2}
\end{equation}
where $\Delta$ is a parameter that determines how sensitive
is the alchemical kernel to differences in
electronegativities. We used $\Delta=1$ to 
generate the last set of maps in Fig.~\ref{fig:qm7b}.
The map now separates out quite accurately regions
with homogeneous stoichiometry. Whereas in the 
$\kappa_{\alpha\beta}=\delta_{\alpha\beta}$ the 
different ``islands'' were roughly arranged according
to a square grid pattern corresponding to $n_\text{O}$ 
and  $n_\text{N}$ along two orthogonal directions, now 
stripe-shaped islands are arranged in 1D, following 
numbers of $n_\text{O}$  and $n_\text{C}$, with 
the number of 
nitrogen atoms coming out clustered in adjacent
``stripes'', but less clear-cut partitioning than
for the other two elements. This is perhaps unsurprising
given that nitrogen has an intermediate electronegativity 
between that of oxygen and carbon, and the metric tries
to separate most efficiently the elements that
differ most based on the alchemical similarity kernel. 

This last example gives perhaps the most compelling
demonstration of how a structural similarity metric
based on a combination of SOAP kernels gives an effective,
broadly applicable and easily customizable strategy
to assess the similarity of materials and molecules, 
and how a sketch-map construction based on such metric
provides an insightful representation of structural
and alchemical landscapes. 

\subsection{Learning molecular properties}

In this paper we focused mainly on the definition of 
a compound structural similarity kernel, and on characterizing
its behavior by means of sketch-map representations. 
It is however important to keep in mind that an 
effective tool to compare atomic structures can find
application to a broad range of problems - one of the
most intriguing being the inexpensive prediction of
physical-chemical properties of materials and molecules.
To demonstrate the great promise of REMatch-SOAP kernels 
for machine-learning of molecular properties, we used
a standard kernel-ridge regression (KRR) method~\cite{hast09book} to reproduce 
the 14 properties that had been reported in Ref.~\cite{qm7b}
for the 7211 molecules we described in the previous paragraph.

We randomly selected 5000 training structures,
and used the remainder as an out-of-sample validation set. 
After having computed the REMatch-SOAP kernel matrix
$\mathbf{K}$ between all the structures, using a cutoff
of 3\AA{} and a regularization parameter $\gamma=0.5$ 
-- in this case including also H atoms in the list of environments --
we computed the KRR weights vector
\begin{equation}
\mathbf{w}=\left(\mathbf{K}_\text{train}^\xi+\sigma \mathbf{1}\right)^{-1} 
\mathbf{y}_\text{train}.
\end{equation}
Here $\mathbf{K}_\text{train}$ and $\mathbf{y}_\text{train}$ are
the kernel matrix and property values restricted to the training set, 
$\xi$ indicates entry-wise exponentiation to tune the spatial range of the kernel, and
$\sigma$ is a regularization hyperparameter. 
The prediction of the properties for the test set can then
be obtained as $\mathbf{y}_\text{test}=\mathbf{K}_\text{test}^\xi \mathbf{w}$, 
where $\mathbf{K}_\text{test}$ is the matrix containing
the REMatch-SOAP kernels between the test points and the training points.
The procedure was repeated 10 times, and the average mean absolute
error (MAE) and root mean square error (RMSE) on the test set
were computed. 

We optimized the $\xi$ and $\sigma$ hyperparameters 
by minimising the MAE on the atomization energy, 
and then used the same values to perform a KRR for all
the other molecular properties. Since we did not further adjust the
choice of kernel and the $\xi$ exponent, all the properties
could be estimated at the same time, as discussed e.g. in Ref.
\cite{rama-vonl15cijc}. 
The results of this procedure are reported in 
Table~\ref{tab:krr-data}, and demonstrate the extraordinary 
performance of REMatch-SOAP for machine-learning applications. 
For the atomization energy we can obtain 
a MAE of less than 1kcal/mol -- 
a four-fold improvement relative to
previous results that were based on a Coulomb matrix
representation of structures and a deep-neural-network 
learning strategy. What is more, even without separately tuning the 
KRR hyperparameters, we can improve or match the 
performance of prior methods for almost all of the 
properties, the only exceptions being some of the 
properties computed with semi-empirical methods.
The fact we can obtain such a dramatic
improvement using a standard regression technique is a 
testament to the effectiveness of our kernel.
The crucial importance of the choice of descriptors
is also apparent by noting that a MAE of about 1.5 kcal/mol was recently obtained by regression based
on a ``bag of bonds'' description of molecules, coupled with a Laplacian kernel~\cite{bag.of.bonds}.

Reaching chemical accuracy in the automated prediction 
of atomization energies is an important milestone, and the
fact that we could achieve that without fully exploring 
the flexibility of the REMatch-SOAP framework (e.g. by optimizing
the entropy regularization parameter, the environment cutoff, 
eliminating the outliers, combining multiple layers
of description or using a non-diagonal alchemical 
similarity matrix) highlights the potential of 
our approach. Future work will be devoted to 
analyzing the performance, convergence and limits of
machine-learning of molecular and materials' properties using
our SOAP-based structural similarity kernel.

\begin{table}[]
\centering
\begin{tabular}{lccccc}
\hline\hline
Property & SD & MAE & RMSE   & MAE\cite{qm7b} & RMSE\cite{qm7b} \\ \hline
$E$ (PBE0) &	9.70  & 0.04 & 0.07	&  0.16 &  0.36  \\ 
$\alpha$ (PBE0) &	1.34 & 0.05 & 0.07	&  0.11 &  0.18  \\
$\alpha$ (SCS) &	1.47 & 0.02 & 0.04	&  0.08 &  0.12  \\
HOMO (GW) &	0.70 & 0.12 & 0.17	&  0.16 &  0.22  \\
HOMO (PBE0) &	0.63 & 0.11 & 0.15	&  0.15 &  0.21  \\
HOMO (ZINDO) &	0.96 & 0.13 & 0.18	&  0.15 &  0.22  \\
LUMO (GW) &	0.48 & 0.12 & 0.17	&  0.13 &  0.21  \\
LUMO (PBE0) &	0.68 & 0.08 & 0.12	&  0.12 &  0.20  \\
LUMO (ZINDO) &	1.31 & 0.10 & 0.15	&  0.11 &  0.18  \\
IP (ZINDO) &	0.96 & 0.19 & 0.28	&  0.17 &  0.26  \\
EA (ZINDO) &	1.41 & 0.13 & 0.18	&  0.11 &  0.18  \\
$E^{*}_{1^{st}}$ (ZINDO) &	1.87 & 0.18 & 0.41	&  0.13 &  0.31  \\
$E_{max}^{*}$ (ZINDO) &	2.82 & 1.56 & 2.16	&  1.06 &  1.76  \\
$I_{max}$ (ZINDO) &	0.22 & 0.08 & 0.12	&  0.07 &  0.12  \\
\hline\hline
\end{tabular}
\caption{Mean absolute errors (MAEs) and root mean square errors (RMSE)
for the KRR estimation of 14 molecular properties, together with previously published estimation~\cite{qm7b} for the same data set.
The standard deviation of the values of the properties across all
7211 molecules in the database is shown in the second column. 
Errors in the KRR estimation refer to a test set of ~2200 randomly selected configurations,
while the remaining structures were used for training. 
 Property labels refer to the level of theory and molecular property, i.e. atomization energy (E), averaged molecular
polarizability ($\alpha$), HOMO and LUMO eigenvalues, ionization potential (IP),
electron affinity (EA), first excitation energy ($E^{∗}_{1^{st}}$), excitation frequency of
maximal absorption ($E^{∗}_{max}$) and the corresponding maximal absorption intensity
($I_{max}$). 
Energies, polarizabilities and intensities are in eV, \AA$^3$
and arbitrary units, respectively.}
\label{tab:krr-data}
\end{table}

\section{Conclusions}

Distances between atomic structures based on combinations
of local similarity kernels provide a flexible framework
to define a metric in structural and alchemical space. 
Atom-centered environment information can be combined
to provide a global measure of (dis)similarity. An average
kernel $\bar{K}$ provides an inexpensive strategy to do so, with 
a cost that scales linearly with the size of the structures
to be compared, but might under-estimate the difference
between two configurations -- since in principle two 
different structures might yield zero $\bar{D}$. 
Alternatively, one can compute the local kernel between
every possible pair of environments (which itself involves
a cost scaling with the square of the number of environments),
and then build a compound kernel $\hat{K}$ by finding the best-match
permutation of the environments -- which gives a
metric with better resolving power, but entails solving a 
cubic-scaling linear assignment problem. 
Introducing an entropy regularization makes it possible
at the same time to reduce the size-scaling to quadratic,
and to obtain a better behaved, smoothly varying 
metric, that interpolates - depending on the regularization
parameter - between the average and best-match limit. 

This strategy to compare atomic configurations builds 
on the very general notion that complex bulk and molecular 
structures arise from the combination of local building 
blocks, and can be applied seamlessly to systems 
as diverse as clusters, bulk phases of an element, 
conformation of a biomolecules and an assembly of 
small chemical compounds with varying atom kinds and 
number. At the same time, the structure of the 
underlying SOAP kernels allows for very effective fine-tuning.
For instance, by choosing the cutoff radius over which 
atomic densities are compared between environments, one
can make the metric more sensitive to the first-neighbor
chemical connectivity, or vice versa, include information
on the long-range conformation of flexible molecules. 
What is more, it is possible to treat structural and 
alchemical complexity on the same footing, by introducing
an alchemical similarity kernel that makes it possible to
specify whether atoms of different species should be 
considered completely separate, or whether a notion of chemical
distance (based e.g. on the difference in 
electronegativity) should be introduced to 
give different weights to substitutions
between elements with similar reactivity. 

We also demonstrate that straightforward application
of the REMatch-SOAP kernel to the ridge-regression 
evaluation of molecular properties matches or outperforms 
 all previously-presented approaches, 
reaching chemical accuracy in the prediction of 
the atomization energies of a set of small organic
molecules. We believe that in this respect we are
only scratching the surface of the potential 
applications to machine-learning of our kernels,
since these results were obtained without using
any of the more sophisticated techniques (e.g. introducing 
a hierarchy of models to capture the variance of properties
 at different structural scales~\cite{hirn+15arxiv}) that have
been shown to significantly improve this kind of 
procedures when using other structural descriptors.

The similarity metric we introduce could find application
as the workhorse of a number of simulation protocols, 
machine-learning algorithms and data mining strategies.
For instance, it could be used to detect outliers in 
automated high-throughput screenings of materials, 
to cluster similar configurations together, to accelerate
the exploration of chemical and conformational space
of materials and molecules. Here, we show in particular
how it can be combined with a non-linear dimensionality
reduction technique such as sketch-map, to give 
simple and insightful two-dimensional representation of
a given molecular or structural data set. As atomistic
modelling adventures into larger-scale structures, and
unsupervised exploration of materials space, 
maps such as these can provide a valuable tool to 
convey intuitive information on complex structural 
and alchemical landscapes, to rationalize 
structure-property relations, and to predict 
physical observables of novel compounds by 
training machine-learning models to libraries 
of known materials.

\section{Acknowledgments}

S.D and M.C would like to acknowledge support from 
the NCCR MARVEL. A.P.B. was supported by a Leverhulme 
Early Career Fellowship with joint funding from 
the Isaac Newton Trust. We would like to thank M. Cuturi
and C. Ortner for insightful discussion. We thank C. Pickard  
for providing silicon structures found with AIRSS, and S. 
Goedecker and M. Amsler for sharing with us the crystal 
structures of low-density silicon polymorphs.

\appendix

\section{SOAP kernel for multi-species environments}

\section{Derivation of the multi-species kernel}
Let us show how the alchemical kernel in Eq.~\eqref{eq:alchemicalKernel} can be derived. The overlap kernel
\begin{equation}
\tilde{k}(\mathcal{X},\mathcal{X}') = \int \mathrm{d}\hat{R} \Biggl | \int \mathrm{d}\mathbf{r} \sum_{\alpha\alpha'} \kappa_{\alpha\alpha'} \rho^\alpha_\mathcal{X} (\mathbf{r})
\rho^{\alpha'}_{\mathcal{X}'} (\hat{R}\mathbf{r}) \Biggl | ^2 
\end{equation}
is first rewritten in terms of the expansion of the atomic density functions ~\eqref{eq:atomicDensitySpecies}
\begin{equation}
\begin{split}
\tilde{k}(\mathcal{X},\mathcal{X}') = \int \mathrm{d}\hat{R} \Biggl |& \sum_{\substack{\alpha \alpha' n\\l m m'}} \kappa_{\alpha\alpha'} \bigl[c^\alpha _{nlm} (\mathcal{X}) \bigr]^\dagger \times \\
&D^l_{mm'}(\hat{R}) c^{\alpha'} _{nlm'} (\mathcal{X'})  \Biggr | ^2
\textrm{,}
\end{split}
\end{equation}
where we carried out the spatial integration. Expanding this result, we obtain
\begin{multline}
\tilde{k}(\mathcal{X},\mathcal{X}') = \int \mathrm{d}\hat{R} \\
 \sum_{\substack{\alpha \alpha' \\ n l m m'}} \kappa_{\alpha\alpha'} \bigl[c^\alpha _{nlm} (\mathcal{X}) \bigr]^\dagger D^l_{mm'}(\hat{R}) c^{\alpha'} _{nlm'} (\mathcal{X'}) \\
 \sum_{\substack{\beta \beta' \\n' l' \\m'' m'''}} \kappa_{\beta\beta'} c^{\beta} _{n'l'm''} (\mathcal{X}) \Bigl[ D^{l'}_{m''m'''}(\hat{R}) c^{\beta'} _{n'l'm'''} (\mathcal{X'}) \Bigr]^\dagger
 \textrm{,}
\end{multline}
which allows us to integrate analytically over all possible rotations $\hat{R}$ and exploit the orthogonality relations of the Wigner rotation matrices.
\begin{multline}
\tilde{k}(\mathcal{X},\mathcal{X}') = \sum_{\substack{\alpha \alpha' n l m m'\\ \beta \beta' n' l' m'' m''}} \frac{8\pi^2}{2 l + 1} \delta_{ll'} \delta_{m m''} \delta_{m'm'''} \times \\ \kappa_{\alpha\alpha'} \bigl[c^\alpha _{nlm} (\mathcal{X}) \bigr]^\dagger c^{\alpha'} _{nlm'} (\mathcal{X'})  \times \\
\kappa_{\beta\beta'}  c^\beta _{n'l'm''}(\mathcal{X}) \bigl[ c^{\beta'} _{n'l'm'''} (\mathcal{X'}) \bigr]^\dagger
\end{multline}
The final formula of the overlap kernel couples the radial, angular, and species channels of the expansion coefficients while being rotationally invariant
\begin{multline}
\tilde{k}(\mathcal{X},\mathcal{X}') = \sum_{\substack{\alpha \alpha' \beta \beta' \\ n n' l m m'}} \frac{8\pi^2}{2 l + 1} \kappa_{\alpha\alpha'} \kappa_{\beta\beta'} \times \\
\bigl[c^\alpha _{nlm} (\mathcal{X}) \bigr]^\dagger c^\beta _{n'lm}(\mathcal{X})
 c^{\alpha'} _{nlm'} (\mathcal{X'})  \bigl[ c^{\beta'} _{n'lm'} (\mathcal{X'}) \bigr]^\dagger
 \textrm{.}
\end{multline}
In terms of the power spectrum~\eqref{eq:pwrMulti} the kernel may be regarded as a dot-product kernel
\begin{equation}
\tilde{k}(\mathcal{X},\mathcal{X}') = \sum_{\substack{\alpha \alpha' \beta \beta'\\n n' l}} p^{\alpha\beta}_{nn'l}(\mathcal{X}) p^{\alpha'\beta'}_{nn'l}(\mathcal{X}') \kappa_{\alpha\alpha'} \kappa_{\beta\beta'}
\textrm{.}
\end{equation}
Finally, it is easy to see that 
one can recover Eq.~\eqref{eq:k-multispecies}
by setting the alchemical kernel to the Kronecker-delta $\kappa_{\alpha\beta} = \delta_{\alpha\beta}$:
\begin{equation}
\tilde{k}(\mathcal{X},\mathcal{X}') = \sum_{\substack{\alpha \beta \\n n' l}} p^{\alpha\beta}_{nn'l}(\mathcal{X}) p^{\alpha\beta}_{nn'l}(\mathcal{X}')
\textrm{.}
\end{equation}

\section{Sinkhorn distance for structural similarity\label{app:sinkhorn}}

Let us discuss briefly how the REMatch procedure can be 
implemented in practice. Consider for generality the $N\times M$
environment similarity matrix $\mathbf{C}(A,B)$ between two 
structures with $N$ and $M$ atoms respectively.
The expression~\eqref{eq:k-regmatch} given in Section~\ref{sec:theory}
for the optimal-transport-inspired
definition of $\hat{K}$
generalizes straightforwardly to non-square matrices~\cite{cutu13nips}:
\begin{equation}
\begin{split}
   &\hat{K}^{\gamma}(A,B) = \operatorname{Tr}\mathbf{P}^{\gamma T}\mathbf{C}(A,B) \\
   &\mathbf{P}^\gamma =\operatorname*{argmin}_{\mathbf{P}\in \mathcal{U}(M,N)} \sum_{ij}   P_{ij} \left(1-C_{ij}+\gamma \ln P_{ij}\right),
   \end{split}
   \label{eq:k-nm}
\end{equation}
where $\mathbf{P}\in \mathcal{U}(N,M)$ is a (scaled) doubly-stochastic
$N\times M$ matrix for which $\sum_i P_{ij}=1/M$ and 
$\sum_j P_{ij}=1/N$.

The Sinkhorn algorithm finds the optimal $\mathbf{P}^\gamma$ by the decomposition 
$\mathbf{P}^\gamma=
\diag{\mathbf{u}}\  \mathbf{K} \diag{\mathbf{v}}=
\mathbf{K}\circ \mathbf{u}\mathbf{v}^T$,
where $\circ$ indicates the Hadamard product, and 
$\mathbf{K}$ is the entry-wise exponential of $(\mathbf{C}-1)/\gamma$, i.e.
\begin{equation}
P^\gamma_{ij} = u_i v_i \exp\left[(C_{ij}-1)/\gamma\right].
\end{equation}
The balancing vectors $\mathbf{u}$ and $\mathbf{v}$ 
can be obtained by the iteration
\begin{equation}
\begin{split}
\mathbf{u}\leftarrow & \mathbf{e}_N/\mathbf{K}\mathbf{v}\\
\mathbf{v}\leftarrow & \mathbf{e}_M/\mathbf{K}^T\mathbf{u}.
\end{split}\label{eq:sinkhorn}
\end{equation}
where $(\mathbf{e}_N)_i=1/N$ are scaled stochastic vectors, 
and the iteration can be initialized by setting 
$\mathbf{v} = \mathbf{e}_M$.

One of the advantages of a regularized match strategy
is that the kernel becomes a smooth function of the 
environment kernels. Computing its derivative $\partial_\alpha K^\gamma$ with 
respect to a parameter $\alpha$ (a Cartesian coordinate, for instance),
is however not completely trivial. 
Such a derivative is in fact
composed of two terms
\begin{equation}
\partial_\alpha\hat{K}^{\gamma}(A,B) = \Tr\mathbf{P}^\gamma\partial_\alpha\mathbf{C}
+\Tr\partial_\alpha\mathbf{P}^\gamma\mathbf{C}.
\end{equation}
The first term is easy to compute -- provided that one can obtain $\partial_\alpha\mathbf{C}$,
the derivative of all environment kernels with respect to $\alpha$. The second term can be further
broken down based on the Sinkhorn decomposition of $\mathbf{P}^\gamma$:
\begin{equation}
\partial_\alpha\mathbf{P}^\gamma = 
\partial_\alpha\mathbf{K}\circ \mathbf{u}\mathbf{v}^T+
\mathbf{K}\circ \partial_\alpha(\mathbf{u}\mathbf{v}^T)
\end{equation}
The critical issue here is that direct evaluation of 
$\partial_\alpha(\mathbf{u}\mathbf{v}^T)$ would involve
performing a separate calculation for each derivative 
$\alpha$, which could make the approach prohibitively 
expensive when, for instance, one would want to 
compute derivatives with respect to the coordinates
of each atom. 

By straightforward albeit tedious algebra, one can 
reformulate the problem in such a way that the 
derivative can be computed cheaply for any
variational parameter, given $\partial_\alpha\mathbf{C}$:
\begin{equation}
\partial_\alpha\hat{K}^{\gamma}(A,B) =
\Tr \mathbf{Q}^T \partial_\alpha\mathbf{C},
\end{equation}
with
\begin{equation}
Q_{ij}=u_{i}K_{ij}v_{j}\left[1+\frac{1}{\gamma}\left(C_{ij}+a_j-Nu_{i}b_i+Mv_{j}c_j\right)\right].
\end{equation}

The $\mathbf{Q}$ matrix can be fully evaluated
based on intermediate terms that do not depend
on $\delta_\alpha\mathbf{C}$:
\begin{equation}
\begin{split}
\mathbf{a}= & -M\mathbf{v}\circ\left(\mathbf{K}\circ\mathbf{C}\right)^{T}\mathbf{u}\\
\mathbf{b}= & \left(\mathbf{1}-\mathbf{W}\right)^{-T}\left[\left(\mathbf{K}\circ\mathbf{C}\right)\mathbf{v}+\mathbf{K}\left(\mathbf{v}\circ\mathbf{a}\right)\right]\\
\mathbf{c}= & N\left(\mathbf{b}\circ\mathbf{u}^{2}\right)\mathbf{K}\\
\mathbf{W}= & \diag \left(N\mathbf{u}^{2}\right)\mathbf{K}\diag\left(M\mathbf{v}^{2}\right)\mathbf{K}^{T},
\end{split}
\end{equation}
were with $\mathbf{u}^2=\mathbf{u}\circ\mathbf{u}$
we indicate the entry-wise square.
The only caveat here is that $(\mathbf{1}-\mathbf{W})$
is singular, and so it cannot be straightforwardly
inverted. Nevertheless, 
$\mathbf{b}$ can be computed
by the fixed-point iteration 
$\mathbf{b}\leftarrow \mathbf{W}^T\mathbf{b}+\mathbf{y}$
with $\mathbf{y}=\left[\left(\mathbf{K}\circ\mathbf{C}\right)\mathbf{v}+\mathbf{K}\left(\mathbf{v}\circ\mathbf{a}\right)\right]$. 
Due to the potential instability of the procedure, it is 
crucial however to check the convergence on the 
overall value of $\partial_\alpha K^\gamma$, and 
not to push the convergence to higher relative accuracy
levels than those used for the original solution to the 
Sinkhorn balancing problem. 

\bibliography{biblio,sandip}

\end{document}